\documentclass[preprint,prd,aps,superscriptaddress,nofootinbib]{revtex4}
\usepackage{graphicx}

\begin{document}

\title {\large VHE $\gamma$ ray absorption by galactic 
interstellar radiation field}

\author{Jian-Li Zhang}
\affiliation{ Key laboratory of particle astrophysics,
Institute of High Energy Physics, Chinese Academy of
Sciences, P.O. Box 918-3, Beijing 100049, P. R. China}

\author{Xiao-Jun Bi }
\affiliation{ Key laboratory of particle astrophysics,
Institute of High Energy Physics, Chinese Academy of
Sciences, P.O. Box 918-3, Beijing 100049, P. R. China}

\author{Hong-Bo Hu}
\affiliation{ Key laboratory of particle astrophysics,
Institute of High Energy Physics, Chinese Academy of
Sciences, P.O. Box 918-3, Beijing 100049, P. R. China}

\date{\today}

\begin{abstract}

Adopting a recent calculation of the Galactic interstellar 
radiation field, we calculate the attenuation of the very
high energy $\gamma$ rays from the Galactic sources. The
infra-red radiation background near the Galactic Center is
very intense due to the new calculation and our result shows
that a cutoff of high energy $\gamma$ ray spectrum begins at
about 20 TeV and reaches about 10\% for 50 TeV $\gamma$ rays.

\end{abstract}

\maketitle

\section{introduction}

The Galactic interstellar radiation field (ISRF) is very important 
in understanding the spectrum of the very high energy (VHE) $\gamma$ rays
originated from the galactic sources. 
At the acceleration site, it is generally 
believed that the VHE cosmic ray electrons interact with ISRF 
photons via inverse Compton scattering and generates the VHE $\gamma$ rays. 
From the propagation's point of view, on the way to the observer, 
the VHE $\gamma$ ray could be absorbed due to its interaction with ISRF 
and in turn changes its spectrum. 

In previous studies the absorption of the VHE $\gamma$ rays is focused on
those from high redshift sources which interact with the extragalactic
photon background. The absorption of the VHE $\gamma$ rays within
the Galaxy is believed to be small since the $\gamma$ rays travel
a short distance. However, due to a most recent calculation by 
Porter and Strong \cite{porter}, which has incorporated a large amount
of relevant new astronomical information of stellar populations, 
Galactic structure and interstellar dust, the density of the infra-red
background within the Galaxy is enhanced considerably compared with
the previous models. Especially there is a very high infra-red
background near the galactic center due to this calculation. Therefore, 
the attenuation of the VHE $\gamma$ rays within the Galaxy may become
important if the $\gamma$ rays come from the direction of the galactic center. 
In another work \cite{allard} %[Porter,Allard2005], 
%ISRF was found to be the major target which significantly decrease 
the mean free path (MFP) of cosmic ray nuclei is calculated again
due to the new results of ISRF. It is found that the MFP is greatly
decreased if considering the infra-red background near the galactic center.
In this paper, we attempt to estimate the absorption 
of VHE $\gamma$ rays within the galaxy by the ISRF.

\section{Opacity of the Galactic ISRF to the VHE $\gamma$ rays}

The ISRF is composed of the star light, the infra-red radiation
and the cosmic microwave background (CMB). The energy density of 
star light is determined by the distribution
and spectrum of each stellar type. 
The starlight is then absorbed by dust in the interstellar
medium and re-emit in infra-red, of which the density depends
on the distribution of the dust.
The starlight is also scattered by the dust and forms 
the diffuse galactic light.
%on the size of the dust grains. 
The starlight dominates the ISRF from 
$0.1\mu m$ to $10\mu m$. The emission from the very small dust 
grain dominates from $10\mu m$ to $30\mu m$ while the emission from 
larger dust grain with temperature of about 20K dominates from
$20\mu m$ to $300\mu m$.
The CMB contributes mainly to above $1000\mu m$. 

A detailed calculation of ISRF has been established and adopted
widely\cite{strong}.
Taking into account the new results of the Galactic stellar 
distribution, the 
dust distribution and other astronomical information of the Galaxy, 
the calculation of the ISRF spectra is recently updated\cite{porter}.
%the spatial dependent ISRF spectra can be estimated and it reproduces 
%the local observed radiation field reasonably well [Moskalenko2000; 
%2004; Porter2005]. 
Together with the very successful infrared surveys carried by the Infrared 
Astronomical Satellite (IRAS) and the Diffuse Infrared Background 
Experiment (DIRBE) on board the Cosmic Ray Background Explorer 
(COBE) satellite,  our knowledge on the 
energy (number) density of ISRF photons has been greatly improved. 
The new calculation of the local radiation field shows quite well 
agreement with these observations. 
%However, owing to the 
%contamination from overwhelming zodiacal light and other 
%background light, the extraction of ISRF from the direct 
%observations heavily relies on the carefully constructed model. 

Due to the ISRF background 
the VHE $\gamma$-rays is attenuated when penetrating the 
radiation background for it undergoes the $e^+e^-$ 
pair production in colliding with the ISRF.
The $\gamma$-ray spectrum $F_0(E)$ at the source is attenuated to 
\begin{equation}
F(E)=F_0(E) e^{-\tau(E,L)}\ ,
\end{equation}
where $F(E)$ represents the observed spectrum after the attenuation,
$\tau(E,L)$ represents the optical depth of the $\gamma$-rays, being
a function of the $\gamma$-ray energy and the distance
from the $\gamma$-ray source. 
%Given the fact that ? ray photon is not deflected by the galactic 
%magnetic field, the calculation of its attenuation e-? attributes 
%to an analytic integration along the line between the source point 
%and the point observer:
$\tau(E,L)$ is given as
\begin{equation}
\tau(E,L)=\int d\cos(\theta) 
\int d E_{\text{ISRF}}
\int_{l.o.s.} dL
\frac{dn(r,E_{\text{ISRF}})}{dE_{\text{ISRF}}}
\sigma_{\gamma\gamma\to e^+e^-}(s)\ \ ,
\end{equation}
where the integration of $dL$ is along the line of sight of the 
incoming $\gamma$-rays, 
$\frac{dn(r,E_{\text{ISRF}})}{dE_{\text{ISRF}}}$ is the number
density of the ISRF photons at the radius $r$ from the GC at energy
$E_{\text{ISRF}}$, $\sigma_{\gamma\gamma\to e^+e^-}(s)$ is the cross
section of the pair production. The cross section is zero below
the threshold when the center-of-mass energy $\sqrt{s} < 2 m_e$.
Above the threshold the cross section is given by
\begin{equation}
\sigma_{\gamma\gamma\to e^+e^-}(s)= \sigma_T\cdot \frac{3m_e^2}{2s}
\left[ -\frac{p}{E}\left( 1+\frac{4m_e^2}{s} \right)
+\left( 1+\frac{4m_e^2}{s} \left( 1-\frac{2m_e^2}{s} \right) \right)
\log\frac{(E+p)^2}{m_e^2} \right]\ \ ,
\end{equation}
where $\sigma_T=8\pi^2\alpha^2/3m_e^2$ is the Thomson cross section
for photon elastic scattering on a rest electron, 
$E=\sqrt{s}/2$ and $p=\sqrt{E^2-m_e^2}$ are the energy and
the magnitude of momentum of the electron at the center-of-mass system.
At the laboratory system, $s$ is given by $s=2 E_\gamma E_{\text{ISRF}}
(1-\cos\theta)$ with $\theta$ being the angle between the momentum 
of the incoming $\gamma$ ray and the ISRF photon.
For the ISRF photon we have assumed an isotropic distribution
when doing the integration over the $\theta$.

%According to the model calculation, the number 
%density of far-infrared ISRF ($20\mu m$ to $300\mu m$) 
%at the Galactic Center 
%(GC) is almost comparable to that of CMB photons.

The attenuation of extragalactic $\gamma$ rays due to 
colliding with the CMB from 
high redshift sources has been studied extensively. 
The attenuation length can fall below 10 Kpc when the $\gamma$ ray
energy reaches about 500 TeV. 
As the new calculation of ISRF shows that the number
density of far-infrared ISRF ($20\mu m$ to $300\mu m$)
at the Galactic Center
is almost comparable to that of CMB photons, we therefore
expect a significant attenuation at the distance of 10 Kpc within
the Galaxy for $\gamma$ rays at the energy of about 50 TeV.
%Similarly to this 
%process, when interacting with far-infrared ISFR, a 50TeV 
%galactic ? rays is expected to be significantly attenuated if 
%the traveling distance reaches10Kpc, in the size of the Galaxy.

\section{Attenuation of the VHE $\gamma$ rays from Sgr A* emission}

\begin{figure}
\includegraphics[scale=0.7]{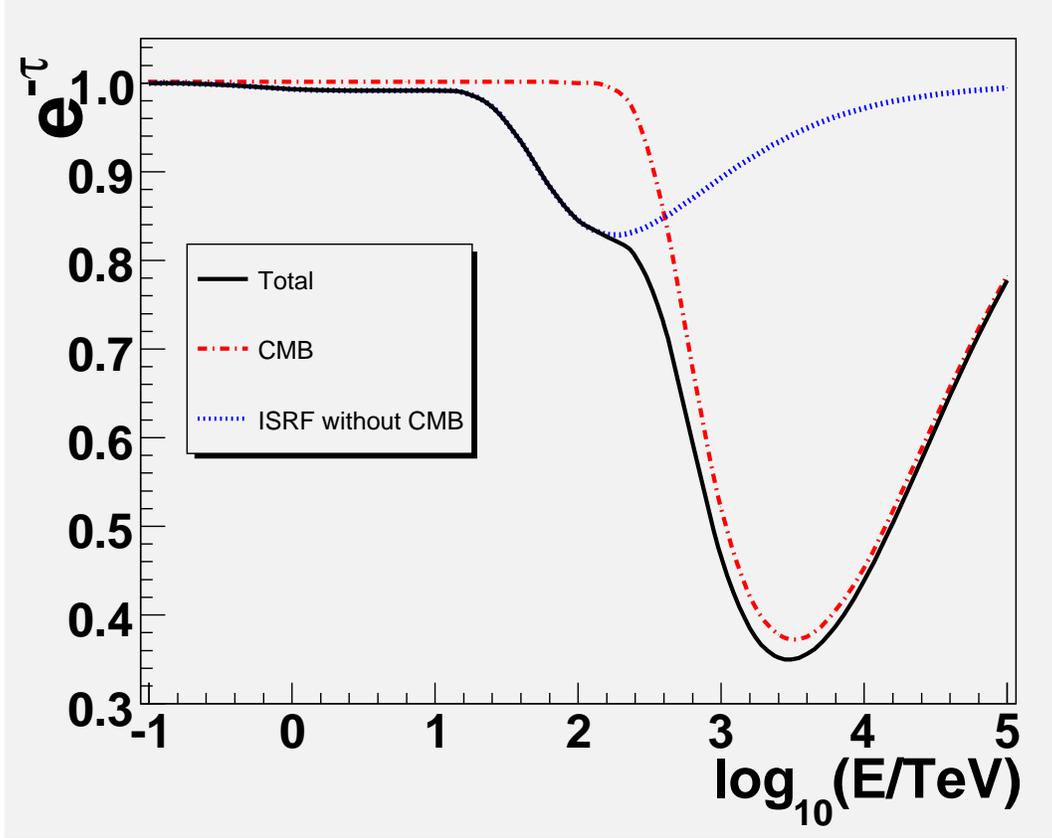}
\caption{\label{atten}
Attenuation of the VHE $\gamma$ rays due to interaction with the
ISRF as a function of the $\gamma$ ray energy. Contributions to the
attenuation due to CMB and the ISRF without CMB are plotted separately.}
\end{figure}

As the ISRF is most intense at the GC, the $\gamma$ rays passing through 
the GC should experience the largest attenuation. Among the limited 
number of TeV $\gamma$ ray sources that have been observed
in the Milky Way, Sgr A* is the
most suitable one for studying this absorption effect. Sgr A* locates 
right at the GC, about 8 Kpc away from the Earth. Though its 
integrated flux above 165 GeV is only 5\% of that of Crab measured
by HESS \cite{hess}, its 
energy spectrum is much harder, with a power law index at about 
2.2 \cite{hess}, %[Aharonian2004], 
which makes its integrated flux above 50 TeV 
almost comparable to that of the Crab and enables future high 
statistic observation.

To calculate the attenuation of the $\gamma$ rays emitted from Sgr A*, 
we need to know the radial distribution of the number densities 
of the ISRF photons between the GC 
and the Earth. As the number density of the ISRF photons falls from the GC 
to the edge of the Galaxy in the form of an exponential function with 
the radial scale length of about 4Kpc \cite{strong,porter},
%[?; Porter,Strong2005],  the  
$dn(r,E_{\text{ISRF}})/dE_{\text{ISRF}}$ 
is therefore interpolated using the densities 
calculated by Porter and Strong\cite{porter} for 
the in-plane at R=0 Kpc, 4 Kpc and 8 Kpc. 
Considering that the dust emission might have a 
different spatial dependence than the overall 
ISRF spectrum, a linear interpolation has also been tried.
Actually there is no difference for the attenuation by
exponential or linear interpolation.

Fig. \ref{atten} shows the calculated attenuation 
for the $\gamma$ rays from Sgr A* 
as a function of $\gamma$ ray energy. 
The cutoff begins at about 20 TeV and about 10\% is abosorbed 
for $\gamma$ rays at 50 TeV and 20\% for $\gamma$ rays at 100 TeV, 
which is  not far from our expectation.
%, 50 TeV $\gamma$ ray is about 10\% abosorbed.
% and it depends upon the detailed  spacial interpolation.
%The attentuation reaches 20\% for $\gamma$ rays at 100 TeV.
The attenuation due to the CMB and the infra-red components
is also plotted to compare their contributions. 
From Fig. \ref{atten} it seems that the effect of VHE
$\gamma$ ray attenuation should show itself if the observation
reaches the energy scale of about 50 TeV.

\section{Discussion}

Using the recently calculated ISRF distribution, 
and taking Sgr A* as an example, this work shows that the ISRF may have
observable attenuation effect to the spectrum of VHE $\gamma$ rays from
the galactic sources. As 
the model might underestimate the ISRF component between 20 to 
40 $\mu m$ compared with the observation by FIRAS\cite{firas}, 
the real effect could be even larger and start to appear 
at even lower energies. In turn, observation on the cutoff energy 
will provide independent information to test and constrain the 
ISRF model.  Recent discovery on the new TeV $\gamma$ 
ray sources on galactic plane \cite{hess2} will provide us more candidates 
to study the distance dependent cutoff effect.

It should be mentioned that even the cutoff of the energy spectrum
is observed the reason can be attributed either to an intrinsic cutoff,
such as the cutoff spectrum of the injecting electron at the $\gamma$
ray source, 
%, or the kinetic constrain in the case of DM annihilation. 
or to the ISRF absorption discussed here. 
Therefore a multi-band observation of the $\gamma$ ray source 
and careful analysis are necessary to reveal the individual
contribution as the intrinsic cutoff  can be traced to lower energy
band while absorption takes place only for the high energy band.
%may indicate such an intrinsic cutoff which
%affects the spectrum of each band associated with this cutoff, while
%the absorption takes place only for the high energy band. 
%The contribution from 
%ISRF can not be distinguished from the other contribution from 
%single source measurement. 
%Proton source is potentially another good candidate in studying 
%the ISRF distribution, as the injecting cosmic ray is proton, the 
%spectrum cutoff should be only originated from ISRF.

\begin{acknowledgments}
This work is supported in part by the NSF of China
under the grant No. 10105004, 10120130794.
\end{acknowledgments}


\begin{thebibliography}{99}

\bibitem{porter}
T.A.Porter and A.W. Strong, astro-ph/0507119.

\bibitem{allard}
T.A.Porter and D. Allard, astro-ph/0507121.
                                                                                
\bibitem{strong}
A.W.Strong, I.V. Moskalenko and O. Reimer, ApJ 537, 763 (2000).

\bibitem{hess}
F. Aharonian et al., [H.E.S.S. Collaboration],
A\& A \textbf{425}, L13 (2004).

\bibitem{firas}
D. Finkbeiner, M. Davis \& D.J.Schlegel, ApJ \textbf{524}, 867 (1999).

\bibitem{hess2}
F. Aharonian et al., [H.E.S.S. Collaboration],
Science 307, 1938 (2005).

\end{thebibliography}
\end{document}